\documentclass[seceq]{ptptex}
\usepackage{wrapft}
\usepackage{latexsym}
\usepackage{graphicx}
\usepackage{amsmath} % need for subequations
\usepackage{wrapft}
\title{Amplification induced by white noise}
\author{
Masamichi \textsc{ishihara}%
\footnote{E-mail: m\_isihar@koriyama-kgc.ac.jp}
}

\inst{%         %Affiliation, neglected when [addenda] or [errata]
Department of Human Life Studies, Koriyama Women's University, \\
Koriyama 963-8503, Japan 
}

\recdate{\today}
\notypesetlogo
\abst{
We investigate the amplification of the field induced by white noise.
In the present study, 
we study a stochastic equation which has two parameters, the energy $\omega(\vec{k})$ of a free particle 
and the coupling strength $D$ between the field and white noise,
where the quantity $\vec{k}$ represents the momentum of a free particle. 
This equation is reduced to the equation with one parameter $\alpha(\vec{k})$ which is defined as 
$\alpha(\vec{k}) = D \left(\omega(\vec{k}) \right)^{-3/2}$.
We obtain the expression of the exponent statistically averaged over the unit time 
and derive an approximate expression of it. 
In addition, the exponent is obtained numerically by solving the stochastic equation. 
We find that the amplification increases with $\alpha(\vec{k})$. 
This indicates that   
white noise can amplify the fields for soft modes if the mass $m$ of the field is sufficiently light and
if the strength of the coupling between the field and white noise is sufficiently strong,
when the energy $\omega(\vec{k})$ is equal to $\sqrt{m^{2} + \vec{k}^{2}}$.
We show that the $\alpha(\vec{k})$ dependence of the exponent statistically averaged is qualitatively similar to 
that of the exponent obtained by solving the stochastic equation numerically,
and that these two exponents for the small value of $\alpha(\vec{k})$ are quantitatively similar. 
}

\begin{document}
\maketitle
\section{Introduction}
Over the past few decades, a considerable number of studies have been made on the phenomena induced by noise
and it is understood that noise plays important roles in many branches of physics. 
Well-known phenomena are stochastic resonance \cite{Gammaitoni,Collins,Yang,Tessone} , 
stochastic synchronization \cite{Chialvo,FUKUDA}, 
noise induced propagation \cite{Zaikin}, 
phase transition induced by multiplicative noise \cite{Broeck},
noise enhanced phase locking \cite{Miyakawa},
coherence resonance \cite{Pikovsky}, 
and so on. 
These examples indicate that noise plays essential roles in some systems.
Therefore, the inclusion of noise is important to explain some physical phenomena. 
A phenomenon that noise may be important 
is the particle production that thermalize the system as discussed 
in the study of the early universe and heavy ion collisions.
The number of produced particles may be enhanced by noise, because noise can supply energy.
Therefore, it is worth to investigate the amplification of the field by noise. 

The amplification of the field has been investigated in terms of particle production,
because the produced particles affect the time evolution of the system. 
An mechanism of the amplification is spinodal decomposition \cite{Felder1} that
the field is amplified through the roll-down from the top of the potential (false vacuum) to 
the minima of the potential (true vacuum), 
because the mass squared is negative around the top of the potential hill.
Another mechanism is parametric resonance 
\cite{Landau,Kofman56,Zanchin,Zanchin2,Ishihara7,Bassett,Muller,Kaiser,Hiro-Oka,Dumitru,Ishihara4,Sornborger,Maedan,Ishihara5}
that an oscillating field amplifies other fields. 
The effects of noise on parametric resonance have been studied \cite{Zanchin, Zanchin2}. 
Noise modifies the amplification quantitatively.
For example, noise shifts enhanced modes \cite{Ishihara7}.

%%%%%%%%%%%%%%%%
These mechanisms are well-known as the mechanism of the amplification of the field through phase transition. 
The amplification by spinodal decomposition occurs in the beginning of the phase transition and 
that by parametric resonance occurs in the end of the phase transition. 
Particle does not produced by parametric resonance if no oscillating field exists, 
because no parametric resonance occurs in the absence of an oscillating field.
However, another mechanism of the amplification may work even in such a case if noise exists. 
%%%%%% INSERTED BEGIN
Phenomenological equations for such processes are similar to the equation of a pendulum.

The oscillator with randomly varying mass and/or friction coefficient has been studied 
in order to comprehend the effects of noise.  
In Ref.~\citen{Stratonovich}, the equations of motion were described with amplitude-angle variables and 
the condition for the growth of amplitude was acquired. 
In the similar manner, the phase transition %, with the magnitude of the amplitude,
was investigated for a pendulum with a randomly vibrating suspension axis \cite{Landa}.
Recently, K. Mallick and P. Marcq investigated the nonlinear oscillator
with white or colored noise \cite{Mallick2002,Mallick2003,Mallick2005eprint}. 
In their studies, the equations of motion were described with energy-angle (or action-angle) variables. 
The power exponents of time development were given by solving the (effective) Fokker-Planck equation 
which describes the evolution of slow variable.
In these studies, the Fokker-Planck equation is used as a main tool. 
%%%%%% INSERTED END

%%%%%%%%%%%%%%%% MODIFIED BEGIN
In this paper, we investigate the effects of noise on the amplification of the fields by  
adopting an equation with white noise. % directly.
We attempt to derive an approximate expression of the exponent and to obtain the exponents
for various values of the parameter.
The role of physical parameters for the amplification 
%(mass, momentum and coupling constant in this paper) 
is obviously shown by the evaluation of the exponents when white noise exists.
%%%%
In \S\ref{sec:amplification}, a basic equation in the present study is introduced. 
The equation describes a harmonic oscillator with varying mass.
An approximate expression of the exponent is derived 
by averaging the term with white noise statistically and by using the steepest descent method, 
without using the Fokker-Planck equation, though the Fokker-Planck equation is a powerful tool. 
%%%%
In \S\ref{sec:numerical_results}, 
we solve the stochastic equation introduced in \S\ref{sec:amplification} numerically. 
Many trajectories can be obtained as numerical solutions and the exponents are extracted by averaging these trajectories.
We compare the values of the exponents obtained from the expression derived in \S\ref{sec:amplification} with 
those of the exponents obtained by solving the stochastic equation. 
%%%% MODIFIED END
Section \ref{sec:conclusion} is devoted to conclusions.

%%%%%%%%%%%%%%%%%%%%%%%%%%%%%%%%%%%%%%%%%%%%%%%%%%%%%%%%%%%
%%%%%%%%%%%%%%%%%%%%%%%%%%%%%%%%%%%%%%%%%%%%%%%%%%%%%%%%%%%
\section{Amplification induced by white noise in a scalar field theory}
\label{sec:amplification}
\subsection{Equation of Motion}
In the beginning of this section, 
we introduce the parameters and the variables used in the following discussion.
The quantity $\omega ( \vec{k} )$ represents the energy with momentum $\vec{k}$. 
$n(t)$ is a white noise which has the properties  
\begin{subequations}
	\begin{align}
\langle n(t) \rangle & =  0, \\
\langle n(t) n(t') \rangle &= \delta(t - t'),
\label{eqn:white_noise_t}
	\end{align}
\end{subequations}
where $\langle \cdots \rangle$ represents the statistical average.
The starting point of this study is the following equation which describes the motion of the field $\phi$
for finite modes near the bottom of the potential with white noise:
\begin{equation}
\frac{d^{2} \phi}{dt^{2}} + \left( \left(\omega(\vec{k})\right)^{2} + D n(t) \right) \phi = 0 , 
\label{eqn:basic1}
\end{equation}
%%%%
where the parameter $D$ is the coupling strength between the noise and the field.
Without loss of generality, we can set this parameter positive or zero.
No oscillating term appears in Eq.~\eqref{eqn:basic1}, 
while an oscillating term appears in the coefficient of $\phi$ in Ref.~\citen{Zanchin, Zanchin2}. 
Equation~(\ref{eqn:basic1}) can describe approximately the system with an oscillating field 
with quite small amplitude or without an oscillating field \cite{Bassett}. 
We regard Eq.~\eqref{eqn:basic1} as a basic equation in this study.
It is not the aim to derive Eq.~\eqref{eqn:basic1} in this paper.
Following discussion is reasonable, once Eq.~\eqref{eqn:basic1} is introduced.

In order to represent Eq.~\eqref{eqn:basic1} by dimensionless variables,
we rewrite Eq.~\eqref{eqn:basic1} in terms of the new variable $z$ given by 
$z=\omega(\vec{k}) \  t$, we obtain 
\begin{equation}
\frac{d^{2} \phi}{dz^{2}} + \left( 1 + \alpha({\vec{k}}) \  r(z;\vec{k}) \right) \phi = 0 ,  
\label{eqn:stochastic_eq}
\end{equation}
where $\alpha(\vec{k})=D \left( \omega(\vec{k}) \right)^{-3/2}$ and 
$r(z;\vec{k}) =  \left( \omega(\vec{k}) \right)^{-1/2} n(z/\omega(\vec{k}))$
which has the property 
\begin{equation}
\langle r(z;\vec{k}) r(z';\vec{k}) \rangle = \delta(z-z').
\label{eqn:mod_noise}
\end{equation}
%%%%%%%%%%%%%%%% Attached on Feb. 16 '06 
The momentum $\vec{k}$ does not play an essential role 
in the calculation of the amplification from Eqs.~\eqref{eqn:stochastic_eq} and \eqref{eqn:mod_noise}
except the momentum dependence of the field amplification.
The amplification is determined by only the value of $\alpha(\vec{k})$. 
%%%%%%%%%%%%%%%%
Equation~\eqref{eqn:stochastic_eq} is rewritten by introducing 
$p_{\phi} = d\phi/dz$ and $dW = r(z;\vec{k}) dz$ , 
because it is not easy to handle Eq.~\eqref{eqn:stochastic_eq} numerically.
It should be noted that $dW$ is a wiener process. 
Therefore, Eq.~\eqref{eqn:stochastic_eq} is rewritten as 
\begin{subequations}
\begin{align}
d \phi &= p_{\phi} \ dz ,  
\label{eqn:p_phi:st1}\\
d p_{\phi} &= - \phi \ dz - \alpha(\vec{k}) \  \phi \circ dW .
\label{eqn:p_phi:st2}
\end{align}
\end{subequations}
We regard Eqs.~\eqref{eqn:p_phi:st1} and \eqref{eqn:p_phi:st2} as stratonovich equations. 
($\circ$ represents a stratonovich equation.)
Stratonovich equation is easily converted to It\^{o} equation. 
The It\^{o} equations corresponding to Eqs.~\eqref{eqn:p_phi:st1} and \eqref{eqn:p_phi:st2} take the same form, that is,  
\begin{subequations}
\begin{align}
d \phi     &= p_{\phi} \ dz ,   
\label{eqn:p_phi:ito1}\\
d p_{\phi} &= - \phi \ dz - \alpha(\vec{k}) \  \phi \ dW .
\label{eqn:p_phi:ito2}
\end{align}
\end{subequations}
We solve Eqs.~\eqref{eqn:p_phi:ito1} and \eqref{eqn:p_phi:ito2} numerically in the next section.

%%%%%%%%%%%%%%%%%%%%%%%%%%%%%%%%%%%%%%%%%%%%%%%%%%%%%%%%%%%%%%%%%%%%%%%%%%%%%%%%%%%%
\subsection{Estimation of the Exponents}
\label{subsec:exponents}
Equation \eqref{eqn:stochastic_eq} describes the motion of a harmonic oscillator 
when $\alpha({\vec{k}})$ is zero. 
Then the solution is a sine function for $\alpha(\vec{k}) = 0 $.
Hereafter, we omit the argument $\vec{k}$ of $\alpha(\vec{k})$ when no confusion occurs. 
The coefficient $\left( 1 + \alpha({\vec{k}}) \  r(z;\vec{k}) \right) $ can be negative for $\alpha \neq 0$, 
because $r(z;\vec{k})$ is a random variable of zero mean value.
($\alpha$ is not negative, because $D$ is larger than or equal to zero.)
In this subsection, we estimate the magnitude of the exponent $\Delta g$
which describes the increase of the amplitude $A(z)$ of the oscillator in time width $\Delta z$:
$A(z + \Delta z) = \exp(\Delta g) A(z)$.
%%%%% ATTACHED PTP
The amplitude does not increase if the coefficient of $\phi$ in Eq.~\eqref{eqn:stochastic_eq} is positive.
Therefore we pay attention only to this coefficient which is negative. 
%%%%%
Then the exponent $\Delta g$ in an extremely short time step $\Delta z$,
in which the random variable $r(z;\vec{k})$ is constant, is given by 
\begin{equation} 
\Delta g = \Delta z \ \Theta \left( - 1 - \alpha({\vec{k}}) \  r(z;\vec{k}) \right)
      \left( - 1 -  \alpha({\vec{k}}) \  r(z;\vec{k}) \right)^{1/2},
\label{eqn:delta_g_def}
\end{equation} 
where $\Theta(x)$ is a step function which is 1 for $x > 0$ and 0 for $x < 0$. 
If the time step $\Delta z$ is large enough compared with the time length 
in which the random variable $r(z;\vec{k})$ is constant, the statistical average of $\Delta g$ should be taken.
In this sense, the width $\Delta z$ is the resolution of the observation. 
To make this fact clear, we divide the time step $\Delta z$ into $N$ regions, in which 
the random noise $r(z;\vec{k})$ is constant.
When the amplification of the field is given by 
$\phi(z_{i} + (\Delta z)/N) = \exp(\beta_{i} (\Delta z)/N) \phi(z_{i-1})$ 
for the region $i$ (The regions are distinguished by the index $i$.), 
the amplification of the field in $\Delta z$ is given by 
\begin{equation}
\phi(z+\Delta z) = \prod_{i=1}^{N} \exp \left( \beta_{i} \ \frac{(\Delta z)}{N} \right) \phi(z)
                 = \exp \left(  \frac{1}{N} \sum_{i=1}^{N} \beta_{i} (\Delta z) \right) \phi(z).
\label{eqn:derive_expression}
\end{equation}
The quantity $\beta_{i} (\Delta z)$ is the exponent in the case that
the random noise is constant in time step $\Delta z$.
Eq.~\eqref{eqn:derive_expression} implies that 
the exponent in $\Delta z$ is the statistical average of the realized exponents $\beta_{i} (\Delta z)$,
when the noise $r(z;\vec{k})$ varies frequently in $\Delta z$.
%%%%% ATTACHED PART PTP
If $\beta_{j}$ is pure imaginary, it does not affect the amplification.
Then step function is included in Eq.~\eqref{eqn:delta_g_def} as stated above.
%%%%% ATTACHED PART END PTP
Therefore we take the statistical average, 
because the Gaussian white noise varies for very small time intervals.
As in the previous subsection, we define $\Delta W = r(z;\vec{k}) \Delta z$. 
The probability distribution function $P(\Delta W)$ of the wiener process $\Delta W$
is a Gaussian distribution of zero mean value and of variance $\Delta z$, that is, 
\begin{equation}
P(\Delta W) = \frac{1}{\sqrt{2 \pi \Delta z}} \exp \left(- \frac{(\Delta W)^{2}}{2\Delta z}\right) .
\label{eqn:distribution_function}
\end{equation}
%%%%%%%%%
We estimate the amplification by $\exp\left[(\Delta g)_{\mathrm st}\right]$, 
where  $(\Delta g)_{\mathrm st}$ is defined by 
\begin{equation}
(\Delta g)_{\mathrm st} = \int_{-\infty}^{\infty} d(\Delta W) P(\Delta W) \Delta g .
\label{eqn:lower_bound}
\end{equation}

%%%%%%%%%%%%%%%%%%%%%%%%%%%%%%%%%%%%
The exponent $(\Delta g)_{\mathrm st}$ with Eq.~\eqref{eqn:distribution_function} can be rewritten as
\begin{equation}
(\Delta g)_{\mathrm st} = 
\frac{\alpha^{1/2} (\Delta z)^{3/4}}{\left(2 \pi\right)^{1/2}} \exp \left(- \frac{\Delta z}{2\alpha^{2}} \right) 
\int_{0}^{\infty} dt \ t^{1/2} \exp \left(-\frac{t^{2}}{2}  - \frac{(\Delta z)^{1/2}}{\alpha} t  \right) .
\label{eqn:gstdz}
\end{equation}
Equation~\eqref{eqn:gstdz} is easily evaluated for $\alpha \gg 1$ and $\alpha \ll 1$.
The exponent $(\Delta g)_{\mathrm st}$ for $\alpha \gg 1$ is given by 
\begin{equation}
(\Delta g)_{\mathrm st} \sim 
\frac{\Gamma(3/4)}{2^{3/4} \pi^{1/2}} \alpha^{1/2} (\Delta z)^{3/4}
\sim 0.411 \alpha^{1/2} (\Delta z)^{3/4} .
\end{equation}
The exponent $(\Delta g)_{\mathrm st}$ for $\alpha \ll 1$ is given by 
\begin{equation}
(\Delta g)_{\mathrm st} \sim \frac{\alpha^{2}}{2^{3/2}} \exp \left(- \frac{\Delta z}{2\alpha^{2}} \right)
\sim 0.354 \alpha^{2} \exp \left( - \frac{\Delta z}{2\alpha^{2}} \right). 
\end{equation}
We introduce an arbitrary integer $n$ which is larger than or equal to 1 
in order to evaluate Eq.~\eqref{eqn:gstdz} approximately for other values of $\alpha$.
Rewriting Eq.~\eqref{eqn:gstdz} in terms of the new variable $x$ given by $t^{1/2} = x^{n}$ and 
using the method of steepest descent, we finally obtain 
%%%%%%%%%%%%%%%%%%
\begin{subequations}
	\begin{align}
(\Delta g)_{\mathrm st} & \sim 
 n \alpha^{1/2} (\Delta z)^{3/4} \left[ f_{n}^{(2)} \left( x_{M}^{(n)} \right) \right]^{-1/2} 
\exp \left( - \frac{\Delta z}{2\alpha^{2}} - f_{n} \left( x_{M}^{(n)} \right) \right)
\nonumber \\  & \quad\quad \times
\mathrm{erfc} \left(- 2^{-1/2} x_{M}^{(n)} \left[ f_{n}^{(2)} \left( x_{M}^{(n)} \right) \right]^{1/2} \right) ,
\label{eqn:steepest_descent}
\\
%%%
& f_{n} \left(x \right) = \frac{1}{2} x^{4n} + \frac{(\Delta z)^{1/2}}{\alpha} x^{2n}  + (1-3n) \ln x, 
\\
& x_{M}^{(n)} = \left\{ 
  \frac{1}{2} \left[ 
  -  \frac{(\Delta z)^{1/2}}{\alpha} 
   + \sqrt{ \left[ \frac{(\Delta z)^{1/2}}{\alpha} \right]^{2} + \left( 6 - \frac{2}{n} \right) } 
               \right] 
         \right\}^{\frac{1}{2n}}
, 
	\end{align}
\end{subequations}
where $f_{n}^{(2)}(x)$ is the second order derivative of  $f_{n}(x)$ with respect to $x$ and 
the error function is defined as
\begin{equation}
\mathrm{erfc}\left(x\right) 
= 
\frac{2}{\sqrt{\pi}} \int_{x}^{\infty} dy \ \exp\left(-y^{2}\right) .
\end{equation}
%%%%% 
For example, 
$(\Delta g)_{\mathrm st}$ from Eq. \eqref{eqn:steepest_descent} for $\Delta z \ll 1$ is approximately 
evaluated as $(\Delta g)_{\mathrm st} \sim 0.428 \alpha^{1/2} (\Delta z)^{3/4}$ when $n$ is set to 1
and $(\Delta g)_{\mathrm st} \sim 0.400 \alpha^{1/2} (\Delta z)^{3/4}$ when $n$ is set to 2.
The quantity $(\Delta g)_{\mathrm st}$ can be evaluated directly from Eq. \eqref{eqn:gstdz} for $\Delta z \ll 1$: 
$(\Delta g)_{\mathrm st} \sim 0.411 \alpha^{1/2} (\Delta z)^{3/4}$.
Then, it is expected that Eq. \eqref{eqn:steepest_descent} gives the approximate value of Eq. \eqref{eqn:gstdz}.
%%%%%%%%%%%%%%%%%%%%%%%%%%%%%%%%%%%%%%%%%%%%%%%%%%%%%%%%%%%%%%%%%%%%%%%%%%%%%%%%%%%%
Our purpose is to evaluate the exponent in the unit time $\Delta z=1$. 
We cannot take $\Delta z \ll 1$, 
because the exponent $(\Delta g)_{\mathrm st}$ is statistically calculated. 
Then, we put $\Delta z =1$ in the following discussions.
In the next section, 
we compare Eq. \eqref{eqn:steepest_descent} with 
the exponent obtained from the numerical solutions for the equation of motion.

%%%%%%%%%%%%%%%%%%%%%%%%%%%%%%%%%%%%%%%%%%%%%%%%%%%%%%%%%%%%%%%%%%%%%%%%%%%%%%%%%%%%
\section{Numerical results}
\label{sec:numerical_results}
In this section, we solve Eqs. \eqref{eqn:p_phi:ito1} and \eqref{eqn:p_phi:ito2} numerically 
for various values of $\alpha$.
The initial conditions are $\phi(0) = 1$ and $\left. d\phi(z)/dz \right|_{z=0} = 0 $ in the present calculations.
We apply the Euler-Maruyama scheme to the stochastic equations.
The time step in $z$ is set to 0.05 in these numerical calculations. 
%%% FIGS 1,2 
\begin{figure}[htb]
        \parbox{\halftext}{%
        \includegraphics[width=\halftext]{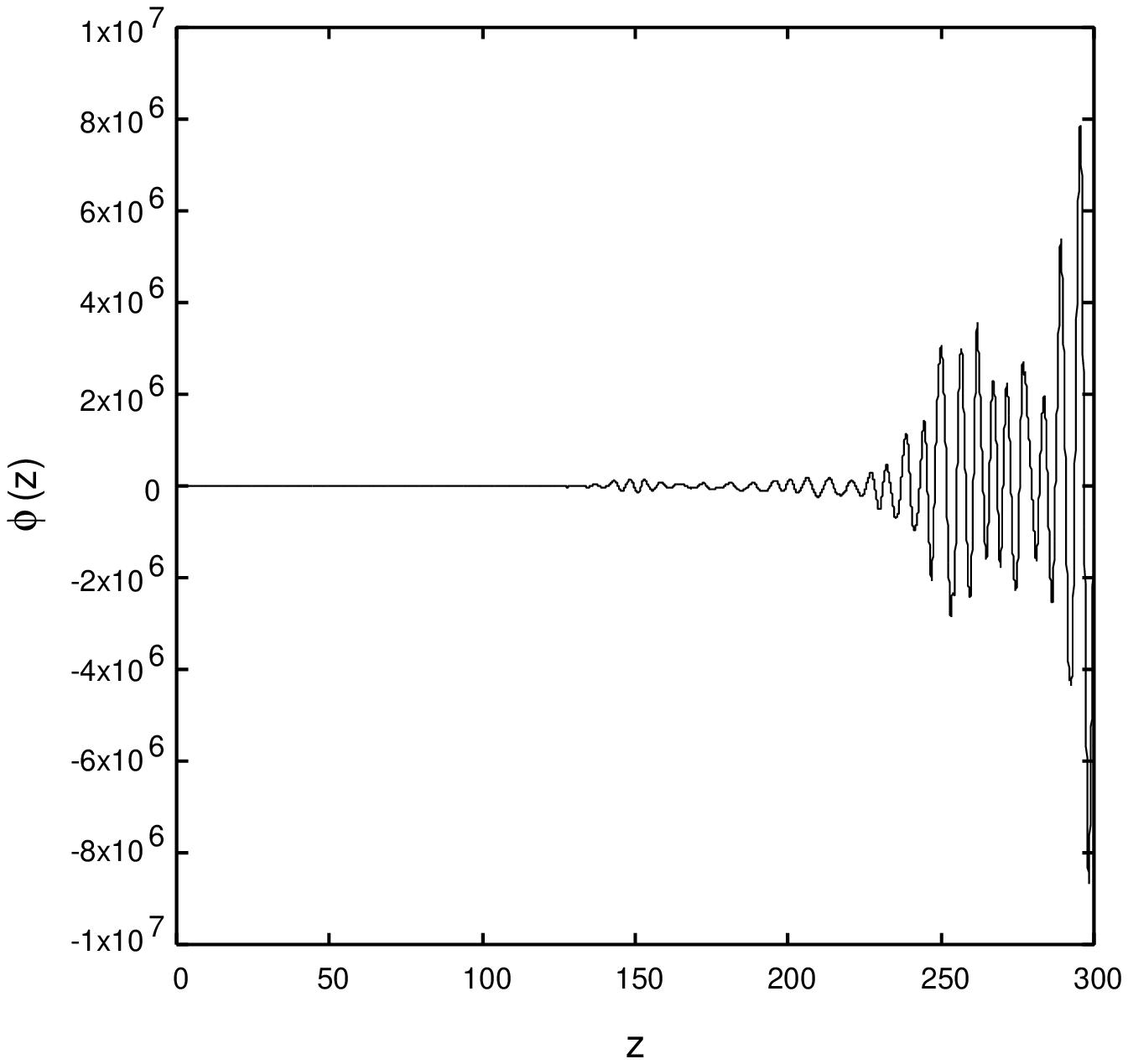}
\caption{
Typical time evolution of the field for $\alpha=0.6$. 
\vspace{1.9cm}
}
	\label{fig:typical_time_evolution}
        }
        \hfill
        \parbox{\halftext}{
        \includegraphics[width=\halftext]{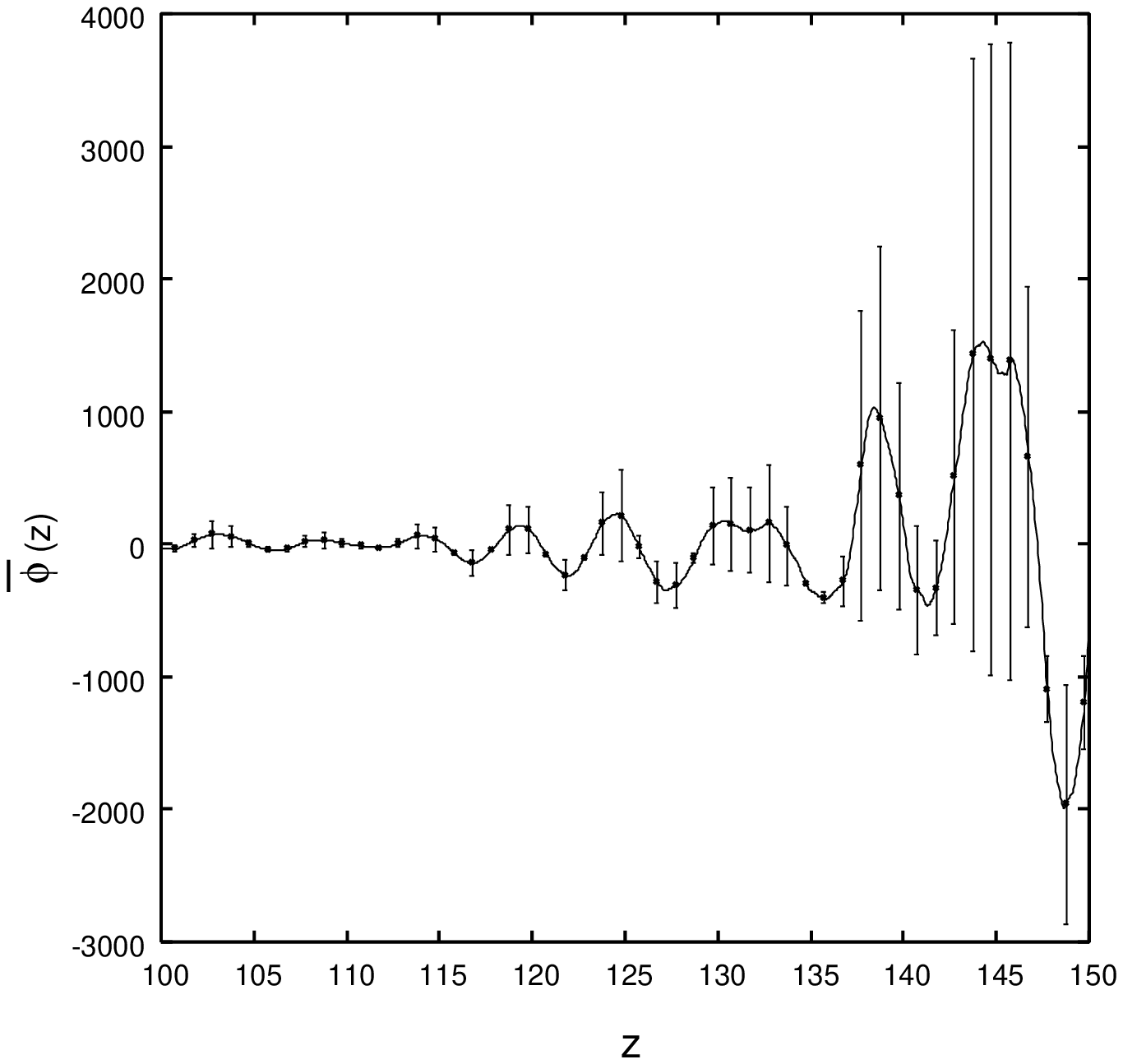}
\caption{
The time evolution of the mean value of $\phi$ for $\alpha = 0.6$. 
One batch has 500 samples and the mean value over 20 batches, $\bar{\phi}(z)$, are calculated numerically . 
Thick curve represents $\bar{\phi}(z)$ and vertical bars indicate the confidential interval of 50\%. 
These bars are depicted every 20 points.
}
	\label{fig:mean}
        }
\end{figure}

%%%%%%%%%%%%%%%%%%%%%
Figure \ref{fig:typical_time_evolution} displays the time evolution of $\phi(z)$ for $\alpha=0.6$. 
Apparently, the field is amplified because of white noise.
One trajectory can be calculated when the sequence of noise is given. 
Thus this amplification may not occur frequently.
Then we should calculate many trajectories and take the average of them. 
%%%%%%%%%%%%%%%%%%%%%
Therefore, we calculate the mean value of the trajectories of the field $\phi_{i}^{(j)}(z)$,
where the subscript $i$ distinguishes different batches and 
the superscript $(j)$ indicates a trajectory in a certain batch. 
One batch has 500 samples (trajectories) and the mean value $M_{i}(z)$ of these trajectories is calculated. 
We take 20 batches and calculate the mean value of $M_{i}(z)$. 
Then, the mean value over 20 batches, $\bar{\phi}(z)$, is given by 
\begin{subequations}
\begin{align}
&\bar{\phi}(z) = \frac{1}{20} \sum_{i=1}^{20} M_{i}(z) , \\
&M_{i}(z) = \frac{1}{500} \sum_{j=1}^{500} \phi_{i}^{(j)}(z) . 
\end{align}
\end{subequations}
The averaged behavior $\bar{\phi}(z)$ for $\alpha=0.6$ from $z=100$ to $150$ is displayed in Fig.~\ref{fig:mean}.
Thick curve represents the mean value $\bar{\phi}(z)$ and 
vertical bars indicate the confidential interval of 50\%. 
These bars are depicted every 20 points in this figure.
These bars become wide with time $z$, 
because the present process is a wiener process.
It is found from this figure that the field is amplified by noise.
We use $\bar{\phi}(z)$ for various values of $\alpha$ in the subsequent evaluation of the exponents.

%%% FIGS 3a,3b 
\begin{figure}[htb]
        \parbox{\halftext}{%
        \includegraphics[width=\halftext]{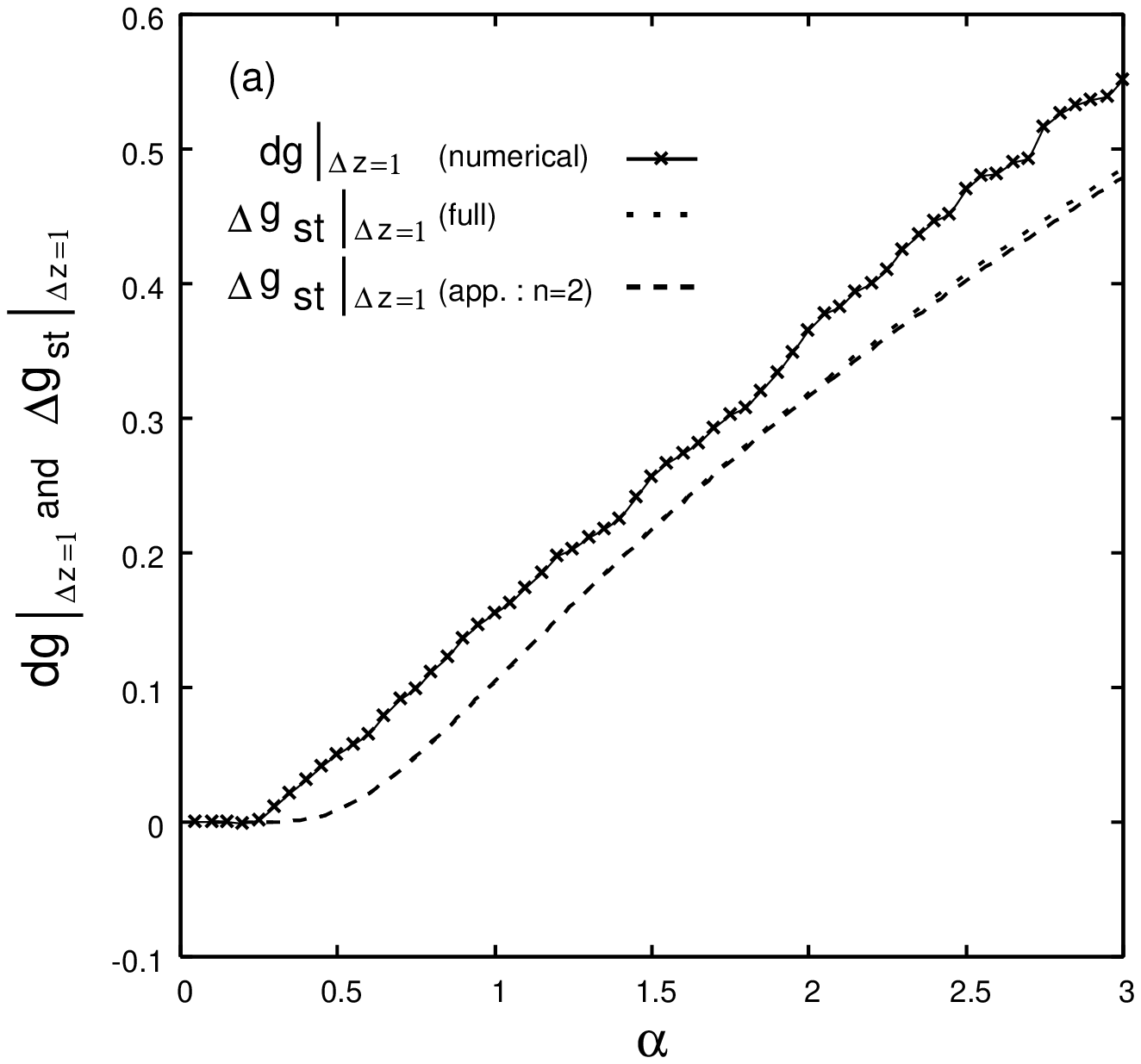}
        }
        \hfill
        \parbox{\halftext}{
        \includegraphics[width=\halftext]{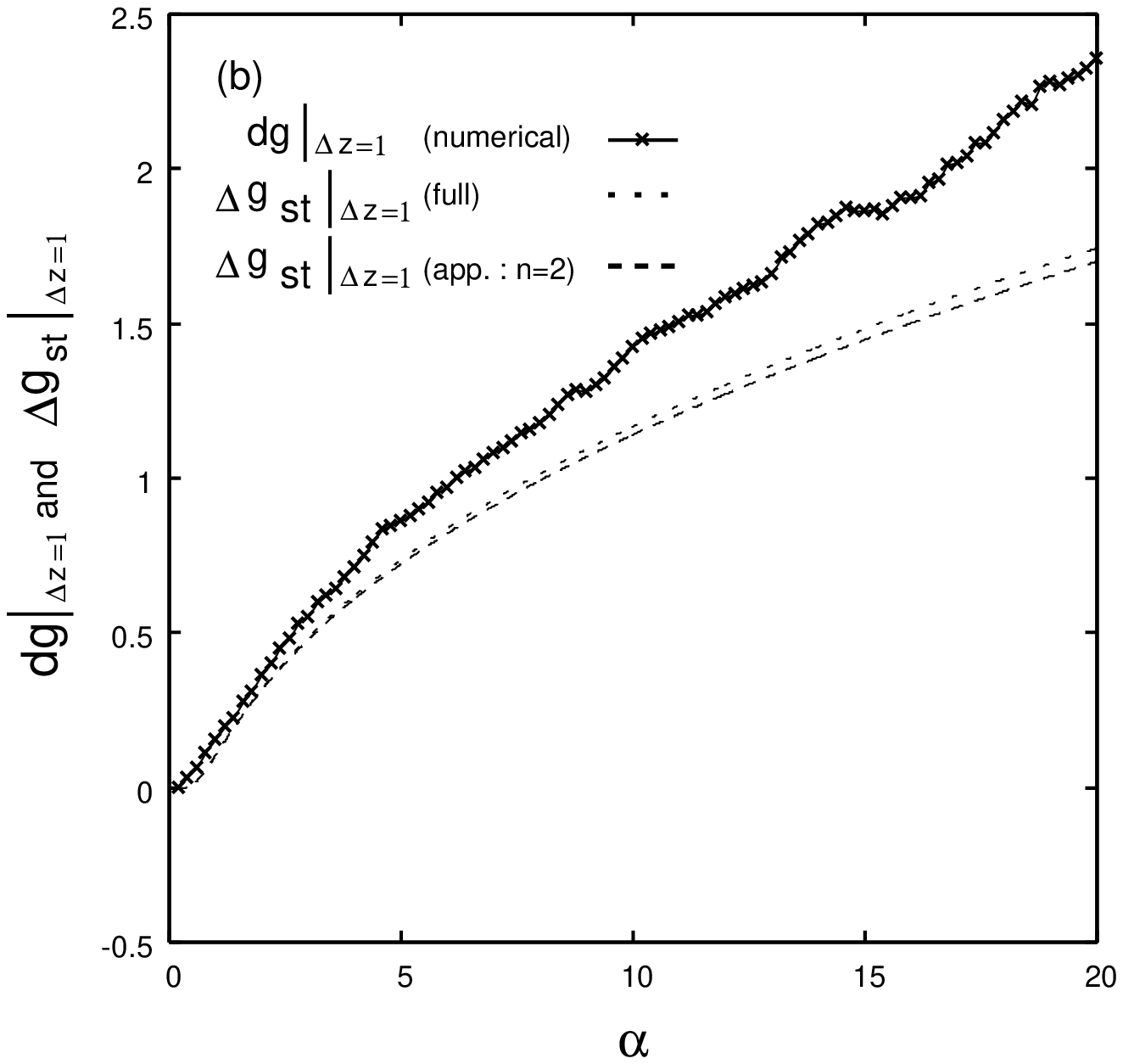}
        }
\caption{
The exponents in the unit time $\left. dg \right|_{\Delta z =1}$ 
and $\left. (\Delta g)_{\mathrm st} \right|_{\Delta z = 1}$, 
in (a) the narrow region ($\alpha  \le 3$) and (b) the wide region ($\alpha  \le 20$). 
The symbol ``$\times$'' represents the data points for $\left. dg \right|_{\Delta z =1}$ 
obtained from $\bar{\phi}(z)$ and the thick curve is a guide.  
The dotted curve represents $\left. (\Delta g)_{\mathrm st} \right|_{\Delta z = 1}$ obtained from Eq.~(\ref{eqn:gstdz})
and
the dashed curve  represents $\left. (\Delta g)_{\mathrm st} \right|_{\Delta z = 1}$ obtained from Eq.~(\ref{eqn:steepest_descent}) with $n=2$. 
}
\label{fig:dgdz}
\end{figure}
%%%%%%%%%%%%%%%%%%%%%%%%%%%%%%%%%%%% Fig. 3a 3b 

%%%%%%
Figure \ref{fig:dgdz} displays the exponents in the unit time
$dg|_{\Delta z = 1}$ and the approximated exponents
in the unit time $\left. (\Delta g)_{\rm st} \right|_{\Delta z = 1}$ for various values of $\alpha$.
The exponent $dg|_{\Delta z = 1}$ is estimated from $\bar{\phi}(z)$ in the region of $z \ge 100$
in order to decrease the effects of the initial conditions. 
This estimation is performed as follows:
1) Making the set of $(z_{i},\ln \bar{\phi}(z_{i}))$, where $z_{i}$ is the time at which 
$\bar{\phi}(z_{i})$ is local maximum and positive.
2) Fitting the above set with a linear function. 
The coefficient of the time $z$ is adopted as $dg|_{\Delta z = 1}$.
The exponent $\left. (\Delta g)_{\mathrm st} \right|_{\Delta z = 1}$ is calculated in two ways. 
One [$\left. (\Delta g)_{\mathrm st} \right|_{\Delta z = 1}$ (full)] is obtained by using Eq. (\ref{eqn:gstdz}) and 
the other [$\left. (\Delta g)_{\mathrm st} \right|_{\Delta z = 1}$ (app. : $n=2$)] is by using Eq. (\ref{eqn:steepest_descent}) with $n=2$.
In Fig.~\ref{fig:dgdz} (a) and Fig.~\ref{fig:dgdz} (b) , 
the symbol ``$\times$'' designates points ``$dg|_{\Delta z = 1}$"  
and the thick curve is a guide. 
The dotted curve represents ``$\left. (\Delta g)_{\mathrm st} \right|_{\Delta z = 1}$ (full)'' and 
the dashed curve represents ``$\left. (\Delta g)_{\mathrm st} \right|_{\Delta z = 1}$ (app. : $n=2$)''. 
In both figures, the dashed and dotted curves are quantitatively similar.
The exponent $dg|_{\Delta z = 1}$ is close to the exponent $\left. (\Delta g)_{\mathrm st} \right|_{\Delta z = 1}$ 
for small values of $\alpha$ as in Fig.~\ref{fig:dgdz} (a) and 
the difference in magnitude between $dg|_{\Delta z = 1}$ 
and $\left. (\Delta g)_{\mathrm st} \right|_{\Delta z = 1}$ increases with $\alpha$ as in Fig.~\ref{fig:dgdz} (b).
Therefore, the exponent $dg|_{\Delta z = 1}$  
can be approximately evaluated with the exponent $\left. (\Delta g)_{\mathrm st} \right|_{\Delta z = 1}$ 
in the region of small values of $\alpha$.
Obviously, numerical results imply that $dg|_{\Delta z = 1}$ 
is a monotone increasing function of $\alpha$. 
In these figures,
the $\alpha$ dependence of $\left. (\Delta g)_{\mathrm st} \right|_{\Delta z = 1}$ 
and that of $dg|_{\Delta z = 1}$ are similar. 
%%%%%%
Note that it is inferred that the exponent $\left. dg \right|_{\Delta z = 1}$ estimated from $\bar{\phi}(z)$ 
is equal to or larger than the exponent $\left. (\Delta g)_{\mathrm st} \right|_{\Delta z = 1}$,
because the arithmetic mean of positive real numbers is equal to or larger than the geometric mean of them. 
When a field $\phi_{j}(z)$ is given by $\phi_{j}(z) =  \exp(\gamma_{j}) \phi(0)$ 
(The index $j$ distinguishes trajectories.) 
and the number of the trajectories is $M$, 
the following inequality is satisfied:
\begin{equation}
|\bar{\phi}(z)| = \frac{1}{M} \sum_{j=1}^{M} \exp(\gamma_{j}) |\phi(0)|  \ge 
\exp \left(\frac{1}{M} \sum_{j=1}^{M} \gamma_{j} \right) |\phi(0)|.
\label{eqn:average_relation}
\end{equation}
%%%%
The exponent of the right-hand side of Eq.~\eqref{eqn:average_relation} is 
the average of exponents obtained from the trajectories.
Assuming that this exponent is equal to the average of the exponents in time (Eq.~\eqref{eqn:lower_bound}),
we can conclude that the exponent $ \left. dg \right|_{\Delta z = 1}$
is larger than or equal to the exponent $\left. (\Delta g)_{\mathrm st} \right|_{\Delta z = 1}$.

%%%%%%%%%%%%%%%%%%%%%%%%%%%%%%%%%%%%%%%%%%%%%%%%%%%%%%%
\section{Conclusions}
\label{sec:conclusion}
We investigate the amplification of the field induced by white noise. 
We obtain the expression of the exponent statistically averaged and 
the expression of the exponent approximated with the method of steepest descent.
In addition, the exponents are extracted from the numerical solutions of the stochastic differential equations. 

We summarize the results: 
%%%
1) White noise amplifies the fields, especially in the region of large values of $\alpha(\vec{k})$. 
   This fact indicates that the amplification for soft modes is strong compared with that for hard modes,  
   when the energy $\omega(\vec{k})$ is equal to $\sqrt{m^{2}+\vec{k}^{2}}$, where $m$ is the mass of the field $\phi$.
   The amplification weakens with the mass of the amplified field. 
%%%
2) The expression of the exponent statistically averaged is reliable to estimate the magnitude of the exponent
   for the small values of $\alpha(\vec{k})$. 
   In particular, the expression approximated with the method of steepest descent is easy to use, 
   because this expression is written with only well-known functions. 
%%%
%%%%%%%%%%%%%%%%%%%%%
These results imply that the soft modes can grow on the vacuum that is located at the bottom of the potential,
if the coupling between the field and white noise (i.e. the parameter $D$) is sufficiently strong
and if the mass of the field $\phi$ is sufficiently light.

%%%%%%
As pointed in \S\ref{subsec:exponents},
the coefficient of $\phi$ in Eq.~\eqref{eqn:stochastic_eq} can be negative because of noise.  
The amplification in the present study is caused by the negative coefficient. 
In comparison, the amplification in spinodal decomposition process is induced by negative mass squared.
Therefore, soft modes are amplified strongly in both the present process and spinodal decomposition process.

It is expected that the stochastic equation used in the present study appears in various fields of physics, 
for example, in the study of the early universe and chiral symmetry rebreaking. 
Though the parameter $\alpha(\vec{k})$ may be one or less, 
the field can be amplified if $\alpha(\vec{k})$ is larger than 0.3. (See Fig.~\ref{fig:dgdz}(a).)
The parameter $\alpha(\vec{k})$ decreases with the mass $m$ 
if the energy $\omega(\vec{k})$ is equal to $\sqrt{m^{2} + \vec{k}^{2}}$.
Therefore, the amplification is strong for shallow potentials.

In the present study, we treat the linear equation with white noise. 
However, nonlinear terms and colored noise appear in general.
We would like to handle the effects of these terms in the future studies.

\end{document}